\documentclass[lettersize,journal]{IEEEtran}

\usepackage[dvips]{graphicx}
\usepackage{latexsym}
\usepackage{epsfig}
\usepackage{color}
\usepackage{amsmath} 
\usepackage{amssymb}
\usepackage{amsxtra}
\usepackage{enumitem}
\usepackage{float}
\usepackage[T1]{fontenc}
\usepackage{setspace}
\usepackage{balance}
\usepackage{diagbox}
\usepackage{epstopdf}
\usepackage{amsthm}
\usepackage{dsfont} 
\usepackage{multirow}
\usepackage{hyperref}
\usepackage{tabularx,colortbl}
\usepackage[table]{xcolor}
\usepackage{lscape}
\usepackage{url}
\usepackage{cite}
\usepackage{algorithm}
\usepackage{algpseudocode}
\usepackage[ subrefformat=parens,labelformat=parens, caption=false, font=footnotesize]{subfig}
\usepackage{mathtools}
\usepackage{units}

\usepackage{flushend}

\usepackage{cases}

\usepackage[cmintegrals]{newtxmath}

\makeatletter
\def\BState{\State\hskip-\ALG@thistlm}
\makeatother

\mathchardef\mhyphen="2D

\newcommand{\abs}[1]{\left|{#1}\right|}

\makeatother

\makeatletter
\def\ps@IEEEtitlepagestyle{%
  \def\@oddfoot{\mycopyrightnotice}%
  \def\@oddhead{\hbox{}\@IEEEheaderstyle\leftmark\hfil\thepage}\relax
  \def\@evenhead{\@IEEEheaderstyle\thepage\hfil\leftmark\hbox{}}\relax
  \def\@evenfoot{}%
}
\makeatother

\def\mycopyrightnotice{%
  \begin{minipage}{\textwidth}
  \centering \scriptsize
    This work has been accepted by the IEEE Signal Processing Letters for publication.  Copyright may be transferred without notice, after which this version may no longer be accessible.
  \end{minipage}
}

\begin{document}

\title{Performance Analysis of Linear Detection under Noise-Dependent Fast-Fading Channels}

\author{
        Almutasem Bellah Enad, \IEEEmembership{Graduate Student Member, IEEE,}
        Jihad Fahs, \IEEEmembership{Member, IEEE,}
        Hadi Sarieddeen, \IEEEmembership{Senior Member, IEEE,}
        Hakim Jemaa, \IEEEmembership{Graduate Student Member, IEEE}
        and Tareq Y. Al-Naffouri, \IEEEmembership{Fellow Member, IEEE}%
            
\thanks{A.~B.~Enad, J.~Fahs, and H.~Sarieddeen are with the American University of Beirut (AUB): aae118@mail.aub.edu, \{jihad.fahs, hadi.sarieddeen\}@aub.edu.lb.
H. Jemaa and T. Y. Al-Naffouri are with King Abdullah University of Science and Technology (KAUST): \{hakim.jemaa, tareq.alnaffouri\}@kaust.edu.sa.
This work is supported by the AUB's University Research Board and Vertically Integrated Projects Program, and KAUST's Office of Sponsored Research under Award No.~ORFS-CRG12-2024-6478.}
}
\maketitle
\begin{abstract}
This paper presents a performance analysis framework for linear detection in fast-fading channels with possibly correlated channel and noise. The framework is both accurate and adaptable, making it well-suited for analyzing a wide range of channel and noise models. As such, it serves as a valuable tool for the design and evaluation of detection algorithms in next-generation wireless communication systems. By characterizing the distribution of the effective noise after zero-forcing filtering, we derive a semi-analytical and asymptotic expression for the symbol error rate under Rayleigh fading and channel-dependent additive circular complex Gaussian noise. The proposed approach demonstrates excellent agreement with integration-based benchmarks as confirmed by numerical simulations thus validating its accuracy. The framework is flexible and can be extended to various channel and noise models, offering a valuable tool for the design and analysis of detection algorithms in next-generation communication systems.
\end{abstract}


\maketitle
\section{Introduction}
\IEEEPARstart{F}{uture} wireless systems must dramatically improve data rates and reliability to support emerging applications~\cite{rajatheva2020white}. This calls for transformative physical layer designs, which combine physics-based models and advanced signal processing for improved spectral and energy efficiency~\cite{Bjornson2024Towards,Sarieddeen9514889}. These developments require a review of conventional performance analysis frameworks to accurately model emerging systems. The symbol error rate (SER) remains a crucial metric for evaluating the reliability of transceiver designs and baseband signal processing~\cite{4544952}, particularly in assessing the robustness of data detection algorithms~\cite{goldsmith2005wireless}. However, to ensure universality of performance analysis, SER derivations must be extended to account for arbitrary channel and noise distributions.

Linear detectors are attractive for their low complexity and near-optimal performance, such as massive multiple-input multiple-output (MIMO) systems under channel hardening conditions~\cite{Yang2015Fifty,8804165,Sarieddeen2024Bridging}. Zero-forcing (ZF) linear detection, in particular, mitigates interference by multiplying the received signals with the channel's inverse (or pseudo-inverse in MIMO). Although in MIMO settings ZF exhibits higher SERs than minimum mean squared error (MMSE) detection at low signal-to-noise ratios (SNR)~\cite{proakis2008digital,se2005fundamentals}, it offers lower computational complexity and does not require knowledge of SNR. ZF SER analysis for modulation schemes such as quadrature amplitude modulation (QAM) has been based on the probability density function (PDF) of the SNR~\cite{4544952,proakis2008digital,se2005fundamentals,258319,Lu1999MPSKAM}, assuming additive Gaussian noise independent of the channel. 
Most studies further assume Gaussian noise after filtering, and closed-form expressions for the effective noise PDFs post-filtering are lacking. Such assumptions lead to {\em overestimating} the probability of error, since the expressions do not take into account the statistical correlation between the channel and the noise, thus not reflecting the reduction in the noise ``uncertainty" given the knowledge of the channel.
In~\cite{pradhan2020ergodic,4533994,dong2019averagesepoptimalprecodingcorrelated}, SER expressions are derived using post-ZF signal-to-interference-plus-noise ratio (SINR) PDFs, but without modeling channel-noise correlation or deriving effective noise statistics.

Deriving the PDF of the post-detection effective noise is crucial for accurate SER analysis in fast fading channels, especially when the noise is channel-dependent due to interference or hardware imperfections~\cite{4544952}. This is also important when detection is followed by noise-centric channel-code decoding~\cite{Sarieddeen2022GRAND}.
As conventional assumptions on channel-noise independence can break down in emerging technologies such as terahertz-band systems and reconfigurable intelligent surface (RIS) links~\cite{Sarieddeen9514889,10663782,9714471,8610080}, our framework accounts for channel–noise dependencies, enabling a more general and accurate analysis framework for next-generation networks.
To model dependencies in wireless communications, Copula theory~\cite{bickel2009copula,nelsen2006introduction} has been widely applied, e.g., for Nakagami-$m$ fading with tail effects~\cite{4487493} and inter-band correlations~\cite{7032317}. Copulas have also been used for MIMO capacity estimation~\cite{Gholizadeh2015OnTC}, multiple-access channel performance under correlated fading \cite{9762969}, and RIS-aided links with phase noise~\cite{9690184}.

In this paper, we consider a single-input single-output (SISO) fast-fading Rayleigh channel with additive Gaussian noise, where the channel and noise terms are, in general, statistically dependent. We develop a generic framework for SER analysis of linear ZF detectors based on the PDF of the filtered effective noise, and we derive corresponding semi-analytical SER expressions and conduct asymptotic analysis. We validate our approach for QAM modulation schemes, demonstrating that our method recovers standard results from the literature when the channel and noise are statistically independent and extends to scenarios with statistical dependence. The dependence is modeled either in the form of a non-zero correlation coefficient or using copula-based techniques~\cite{nelsen2006introduction}. We validate our approach with consistent numerical results. Our proposed framework can be extended to optical systems~\cite{Elsayed2024}.

\section{Problem Formulation}
\label{sec:sysmodel}
We consider a narrowband SISO communication system with a frequency-domain input-output relationship, $r = hs + n$, where $r$ denotes the received symbol, $s$ the transmitted QAM symbol, $h$ the channel coefficient, and $n$ the additive noise. We assume that $h$ and $n$ are jointly Gaussian, with each being a complex circular Gaussian variable with zero mean and respective variances $\mathbb{E}[hh^*] = \sigma_h^2$ and $\mathbb{E}[nn^*] = \sigma_n^2$. 
At the receiver, a ZF detector inverts the channel, $\tilde{r} = h^{-1}r = s+\frac{n}{h} = s + Z$,
where $Z$ denotes the effective noise. 
We consider the maximum likelihood (ML) detector.
Finding optimal likelihood-ratio thresholds is complex for non-Gaussian noise \cite{Kassam1988}. However, the post-filtering decision regions of the ML detector remain Gaussian under perfect channel state information (CSI) at the receiver. Thus, for detection, the effective noise is Gaussian with variance \(\tilde{\sigma}^2 \!=\! \sigma^2/|h|^2\) (for a given $h$). 
We consider $M$-QAM constellation sets with equal a priori probabilities, $1/M$. The ML decision boundaries between each pair of symbols, \((s_m, s_n)\), for \(m, n = 1, 2, \cdots, M\) and \(m \neq n\) are determined by the minimum distance rule.
The average SER is expressed as
\begin{align}
    &\text{SER}= 1 - \frac{1}{M}\sum_{k=1}^{M} P_{c|s_k},\label{PE-16QAM}
\end{align}
where $P_{c|s_k}$ is the probability of correct detection of transmitted symbol \(s_k = (\gamma_k,\beta_k)\),
\begin{equation}
    P_{c|s_k}= \hspace{-1pt} 1-P_{e|s_k}  =\iint_{D_{k}} f_{Z_r,Z_i}(x-\gamma_k, y - \beta_k)\, dx\, dy,\label{eq:gen}
\end{equation}
where $x$ and $y$ denote the real and imaginary components of the received signal $r$, $D_k$ is the corresponding decision region,
 and $f_{Z_r,Z_i}(\cdot,\cdot)$ is the PDF of the effective noise.

\section{Reference SER Analyses Framework}\label{sec:ref-an-fr}
This section reviews related work on SER analysis under the assumption of noise-channel independence. The symbol error probability for $M$-QAM modulation over a Rayleigh fading channel is derived in closed-form~\cite[eq. 8.107]{4544952}.
A general approach to derive the ZF SER under perfect CSI is to 
assume ergodicity~\cite{4544952,proakis2008digital}. 
For a transmitted symbol $s_1 = (\gamma_1,\beta_1)$, 
\begin{equation}
    P_{c|s_1} \hspace{-1pt}=\hspace{-1pt} 1-P_{e|s_1} \hspace{-1pt}=\hspace{-1pt} \iint_{D_1} \hspace{-1pt}\mathbb{E}_h\left[f_N\left((\gamma_1,\beta_1),\frac{\sigma_n^2}{|h|^2}\right)\right]\hspace{-1pt}dxdy,
    \label{eq:stan}
\end{equation}
 where \(f_N\left((\gamma_1,\beta_1),\frac{\sigma_n^2}{|h|^2}\right)\) denotes the PDF of a Gaussian distribution with mean \((\gamma_1,\beta_1)\) and covariance matrix $(\sigma_n^2/|h|^2)\mathbb{I}_2$, and $D_1$ is the decision region corresponding to $s_1$.

When $\abs{h}$ and $\abs{n}$ are Rayleigh distributed with variances $\sigma_h^2$ and $\sigma_n^2$, respectively, the channel and noise PDFs are
\begin{equation}
    f_{|h|}(\psi) = \frac{\psi}{\sigma_h^2} \exp\left( -\frac{\psi^2}{2\sigma_h^2} \right),\ \ 
    f_{|n|}(\nu) = \frac{\nu}{\sigma_n^2} \exp\left( -\frac{\nu^2}{2\sigma_n^2} \right).\label{channel_and_noise}
\end{equation}
From~\cite[3.461.2b]{gradshteyn2014table}, 
\begin{align}  
\mathbb{E}_h&\left[f_N\left(\left(\gamma_1,\beta_1\right),\frac{\sigma_n^2}{|h|^2}\right)\right]=\nonumber\\ &= \int_0^\infty \hspace{-2.4mm}\frac{\psi^3}{2\pi\sigma^2}  \exp\left(-\psi^2\left(\frac{(x-\gamma_1)^2 \!+\! (y-\beta_1)^2\!+\!\sigma^2}{2\sigma^2} \right) \right)\! d\psi
\nonumber\\ &=\frac{1}{\pi} \frac{\sigma^2}{(\sigma^2 + (x-\gamma_1)^2 + (y-\beta_1)^2)^2},
\end{align}
where we assumed, without loss of generality (WLoG), \(\sigma_h^2 = 1\) and \(\sigma_n^2 = \sigma^2\). Therefore, the average SER given $s_1$ is
\begin{equation}
    P_{e|s_1}=1-\iint_{D_1} \frac{1}{\pi} \frac{\sigma^2}{(\sigma^2 + (x-\gamma_1)^2 + (y-\beta_1)^2)^2} dxdy. \label{Pc-eq ref}
\end{equation} 
The same procedure applies to other constellation symbols.

\section{Proposed SER Analyses Framework}
\label{sec:stat-ana}
In this section, we present an alternative method for computing the SER, based on the statistics of the effective noise defined as the ratio between the additive noise and the channel fading, which are not necessarily statistically independent.  

\subsection{Statistical Correlation Model}\label{sec3-1}
 In this setup, effective noise is the ratio between two correlated complex Gaussian terms, $Z = \frac{n}{h}$. 
The PDF of $Z$ is given by~\cite{8839853},
\begin{equation}
    f_{Z_r,Z_i}(z_r,z_i)=\frac{1 - \abs{\lambda}^2}{\pi \sigma_n^2\sigma_h^2} 
    \left( \frac{\abs{z}^2}{\sigma_n^2} + \frac{1}{\sigma_h^2} - 2\frac{\lambda_r z_r - \lambda_i z_i}{\sigma_h\sigma_n} \right)^{-2}, \label{pdf-corr}
\end{equation}
where \(\lambda \!=\! \lambda_r + j\lambda_i\) is the complex coefficient of correlation between $n$ and $h$. Setting \(\sigma_h^2 \!=\! 1\) and \(\sigma_n^2/\sigma_h^2\!=\!\sigma^2\), WLoG, we use~\eqref{eq:gen} and the expression in~\eqref{pdf-corr} to find the average SER given that symbol $s_1$ is transmitted,
\begin{align}
    P_{e|s_1} =&\left(\frac{1 - \abs{\lambda}^2}{\pi\sigma^2}\right)\sum_{k=2}^M \iint_{D_k}
    \left( \frac{(x-\gamma_1)^2 + (y-\beta_1)^2}{\sigma^2} + 1 \right. \notag \\
    &\left. \qquad \quad - 2\cdot\frac{\lambda_r(x-\gamma_1) - \lambda_i(y-\beta_1)}{\sigma} \right)^{-2} \, dx \, dy. \label{Pc-eq-corr}
\end{align}
Equation~\eqref{Pc-eq-corr} provides a semi-analytical expression of the probability of detection error under a potential noise-channel correlation. The independent case captured by~\eqref{Pc-eq ref} is recovered by inserting $\lambda = 0$. In fact, it can be shown that the proposed~\eqref{eq:gen} and reference~\eqref{eq:stan} methods are equivalent when $n$ and $h$ are independent, regardless of the probability distributions under study. 
\subsubsection*{Asymptotic Analysis}
To simplify the analysis and gain insights, we conduct an asymptotic analysis of~\eqref{Pc-eq-corr} at high SNR for $M=4$ and when the $\{D_k\}$s, $1 \leq k \leq 4$, are determined by the four quadrants. Whenever $\sigma\rightarrow 0$, the behavior of the integral in~\eqref{Pc-eq-corr} is primarily dictated by the term that scales inversely with the square of the noise variance. 
Therefore, it is sufficient to asymptotically approximate the integral by retaining only the dominant quadratic term.
Hence, Equation~\eqref{Pc-eq-corr} boils down to
{\small\begin{align}
    P_{e|s_1} &=\Lambda \sum_{k=2}^4\iint_{D_k}\frac{ dx \, dy}{((x-\gamma_1)^2 + (y-\beta_1)^2)^{2}}= \Lambda \sum_{k=2}^4 I_{D_k}, \label{SER_asym}
\end{align}}
where $\Lambda \triangleq \frac{\sigma^2 (1 - \lambda^2)}{\pi}$. Starting from $D_2=[-\infty,0]\times[0,\infty]$, we apply $u = x - \beta_1$ and $v = y - \gamma_1$, to get
\begin{align}
    I_{D_2} = \int^{-\gamma_1}_{-\infty}\int_{-\beta_1}^\infty\frac{du\, dv}{(u^2 + v^2)^{2}}.\label{EQ_par}
\end{align}
Considering the inner integral, we have
\begin{align}
    I &= \int_{-\beta_1}^\infty\frac{du}{(u^2 + v^2)^{2}}\notag \stackrel{(\mathrm{a})}{=} -\frac{1}{v^3}\int_{\arctan\left(\frac{\beta_1}{v}\right)}^{\frac{\pi}{2}} \cos^2{(\theta)}  d\theta \notag\\
    & = \frac{-1}{2v^3}\!\left[\theta\!+\!\frac{\sin{2\theta}}{2}\right]_{\arctan\left(\frac{\beta_1}{v}\right)}^{\frac{\pi}{2}} \!\!\!\!\stackrel{(\mathrm{b})}{=}\!\!\frac{1}{2v^3}\!\left[-\frac{\pi}{2}\!+\!\arctan
    \left(\frac{\beta_1}{v}\right)\!+\!\frac{\beta_1 v}{v^2\!+\! \beta_1^2} \right]. \label{eq_arctan}
\end{align}
In step (a), we apply the change of variable \( u \!=\! -v \tan(\theta) \), 
 and in step (b), we use the trigonometric identity \( \sin(2\theta) = \frac{2\tan(\theta)}{1+\tan^2(\theta)} \). 
 Equation~\eqref{EQ_par} becomes
\begin{align}
I_{D_2} &= \frac{1}{2} \int^{-\gamma_1}_{-\infty}\left[-\frac{\pi}{2v^3}+\frac{\arctan\left(\frac{\beta_1}{v}\right)}{v^3}+\frac{\beta_1}{v^2(v^2+ \beta_1^2)} \right]dv \nonumber\\
 &=\frac{\pi}{8\gamma_1^2}+\underbrace{\int^{-\gamma_1}_{-\infty}\frac{\arctan\left(\frac{\beta_1}{v}\right)}{2v^3} dv}_{I_1 \, \text{integral}}+\underbrace{\int^{-\gamma_1}_{-\infty}\frac{\beta_1}{2v^2(v^2+ \beta_1^2)}\,dv}_{I_2 \,\text{integral}}. \notag
\end{align}
We start with $I_1$:
\begin{align}
I_1     &\stackrel{(\mathrm{a})}{=} \frac{1}{2\beta_1^2}\int_{-\frac{\beta_1}{\gamma_1}}^{0}t\,\arctan(t)\,dt\notag\\
        &\stackrel{(\mathrm{b})}{=}\frac{1}{4\beta_1^2}\left[\left[t^2\arctan(t)\right]_{-\frac{\beta_1}{\gamma_1}}^{0} -\int_{-\frac{\beta_1}{\gamma_1}}^{0} \frac{t^2}{1+t^2}\,dt  \right]\notag\\
        &=\frac{1}{4\beta_1^2}\left[t^2\arctan(t)-t+\arctan(t)\right]_{-\frac{\beta_1}{\gamma_1}}^{0}\notag\\
        &=\frac{\arctan(\frac{\beta_1}{\gamma_1})}{4\gamma_1^2}+\frac{\arctan(\frac{\beta_1}{\gamma_1})}{4\beta_1^2}-\frac{1}{4\gamma_1\beta_1}, \label{I_1_integral}
\end{align}
where in step (a), we perform a change of variable by setting \( t \!=\! \frac{\beta_1}{u} \), and 
in step (b), we perform integration by parts. 
Regarding $I_2$:
\begin{align}
I_2&\stackrel{(\mathrm{a})}{=}\frac{1}{2\beta_1}\int^{-\gamma_1}_{-\infty}\left[\frac{1}{u^2}-\frac{1}{(u^2+ \beta_1^2)} \right]\notag\\
        &=\frac{1}{2\beta_1\gamma_1}+\frac{\arctan\left(\frac{\gamma_1}{\beta_1}\right)}{2\beta_1^2}-\frac{\pi}{4\beta_1^2}, 
\end{align}
where step (a) is due to a partial fraction expansion. 
Because of the geometric symmetry of the integration domains, $I_{D_2} = I_{D_4}$. Furthermore, we compute $I_{D_3}$ following a similar approach.  
 Finally, the asymptotic expression of the average SER is
\begin{align}
&P_{\mathrm{e|s_1}} \notag\\ &\!\approx \!\Lambda \!\left[ 
   \!\frac{1}{4\gamma_1\beta_1}\! +\! \frac{3\pi}{8\gamma_1^2}\! -\! \frac{\pi}{4\beta_1^2}\!+ \! \left(\frac{\gamma_1^2 \!+\! \beta_1^2}{4\gamma_1^2\beta_1^2}\right) \arctan\left(\frac{\beta_1}{\gamma_1}\right)
   \!+\! \frac{\arctan\left(\frac{\gamma_1}{\beta_1}\right)}{2\beta_1^2}
\right]\!.
\label{SER_corr}
\end{align}
Equation~\eqref{SER_corr} captures the effect of statistical correlation on the SER through the multiplicative term, $\Lambda=\frac{\sigma^2(1 - \abs{\lambda}^2)}{\pi}$. As $\lambda$ increases, the SER decreases, a behavior that is caused by the fact that knowing the channel under correlation reduces the uncertainty about the noise, thus yielding lower SER values. According to equation~\eqref{SER_corr}, the channel-noise correlation offers a gain equal to $10 \log_{10}(1 - |\lambda|^2)$ dB, which we call the {\em correlation gain}. 

Naturally, equation~\eqref{SER_corr} is valid for the independent case as well, simply by plugging in $\lambda = 0$ in $\Lambda$. However, we next conduct a more accurate approximation of the $\lambda = 0$ case, one that holds for a wider range of SNR values.
We start by considering equation~\eqref{Pc-eq-corr} for $\lambda = 0$. We begin with $D_2$ and perform the change of variables $u = x - \beta_1$ and $v = y - \gamma_1$:
\begin{align}
&P_{e|(s_1,D_2)} \notag\\
&= \frac{\sigma^2}{\pi}\int^{-\gamma_1}_{-\infty}\int_{-\beta_1}^\infty  \frac{1}{(\sigma^2 + u^2 + v^2)^2} du\,dv \notag\\
&= \frac{\sigma^2}{\pi}\int^{-\gamma_1}_{-\infty} \frac{1}{2a^3}\left[ \frac{\pi}{2}+\arctan\left(\frac{\beta_1}{a}\right)+\frac{\beta_1 a}{a^2+ \beta_1^2}\right]\, dv, \label{eq:sameasbefore}
\end{align}
where $a \triangleq \sqrt{\sigma^2+v^2}$, and where equation~\eqref{eq:sameasbefore} is justified using the same steps that lead to~\eqref{eq_arctan}. To solve equation~\eqref{eq:sameasbefore}, we set:
\begin{align}
    &I_3=\int^{-\gamma_1}_{-\infty}\frac{\pi}{2(\sigma^2 + v^2)^{\frac{3}{2}}} \,dv, \,I_5=\int^{-\gamma_1}_{-\infty}\frac{\arctan\left(\frac{\beta_1}{\sqrt{\sigma^2 + v^2}}\right)}{(\sigma^2 + v^2)^{\frac{3}{2}}}\,dv  \notag\\
    &I_4= \int^{-\gamma_1}_{-\infty}\frac{\beta_1}{(\sigma^2 + v^2)(\sigma^2 + v^2+\beta_1^2)}\,dv. \notag
\end{align}
Starting with $I_3$:
\begin{align}
    I_3&\stackrel{(\mathrm{a})}{=}\frac{\pi}{2 \sigma^2}\int^{\arctan\left(\frac{-\gamma_1}{\sigma}\right)}_{-\frac{\pi}{2}}\cos \theta\, d\theta
    &\stackrel{(\mathrm{b})}{=}\frac{\pi}{2 \sigma^2} \left( 1 - \frac{\gamma_1}{\sqrt{\sigma^2 + \gamma_1^2}} \right),
\end{align}
 where in step (a) we used \( v = \sigma \tan(\theta) \), 
 and in step (b), we use the trigonometric identity, 
  $\sin(\theta) = \frac{x}{\sqrt{1+x^2}}$, \text{where} $x = \arctan(\theta)$. 
As for $I_4$: 
\begin{align}
    I_4
    &\stackrel{(\mathrm{a})}{=}\frac{1}{\beta_1}\int^{-\gamma_1}_{-\infty}\left[\frac{1}{(\sigma^2 + v^2)}-\frac{1}{(\sigma^2 + v^2+\beta_1^2)}\right]\,dv \notag\\
    &= \frac{1}{\beta_1} \left[\frac{\frac{\pi}{2}-\arctan(\frac{\gamma_1}{\sigma})}{\sigma}+\frac{ \arctan(\frac{\gamma_1}{\sqrt{\sigma^2+\beta_1^2}})-\frac{\pi}{2} }{\sqrt{\sigma^2+\beta_1^2}}\right],
\end{align}
where we used partial fraction expansion in step (a). 
We approximate \( I_5 \) as $\sigma \rightarrow 0$ to get
\begin{align}
    I_5 \approx \int^{-\gamma_1}_{-\infty}\frac{\arctan\left(\frac{\beta_1}{{v}}\right)}{v^3}\,dv. 
\end{align}
As previously demonstrated in~\eqref{I_1_integral}, we find $I_5$ as  
\begin{align}
I_5=\frac{\arctan\left(\frac{\beta_1}{\gamma_1}\right)}{2\gamma_1^2}+\frac{\arctan\left(\frac{\beta_1}{\gamma_1}\right)}{2\beta_1^2}-\frac{1}{2\gamma_1\beta_1} .
\end{align}
\begin{figure*}[ht!]
\vspace{-8mm}
 \centering
 \subfloat[Independent channel and noise.]{\label{fig-ind} \includegraphics[width=0.32\linewidth]{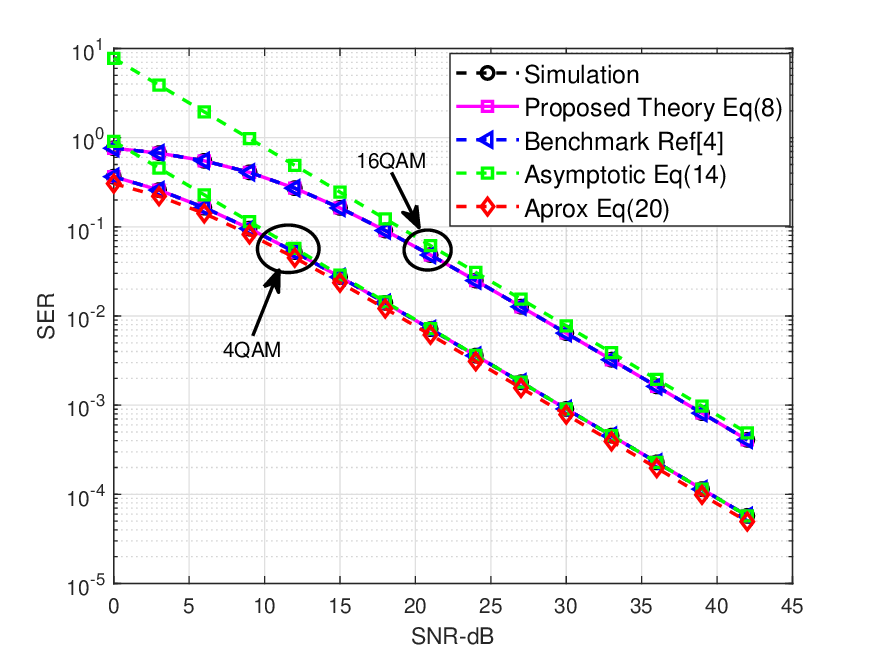}}%
 \hfill
\subfloat[Correlated channel and noise using the statistical correlation model.]{\label{fig-corr} \includegraphics[width=0.32\linewidth]{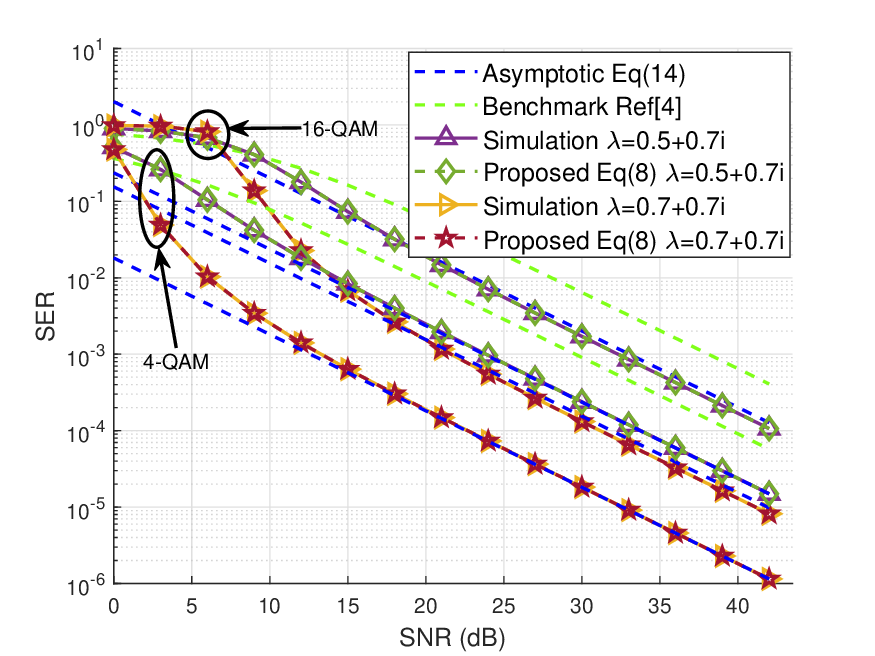}}%
 \hfill
 \subfloat[
 Correlation using Copula method~\eqref{eq:frank}. Theory refers to eq.~\eqref{PE-16QAM} with $f_{\Re(z), \Im(z)}(\cdot,\cdot)$ given by~\eqref{effectiv-copula}.]{\label{fig-frank} \includegraphics[width=0.32\linewidth]{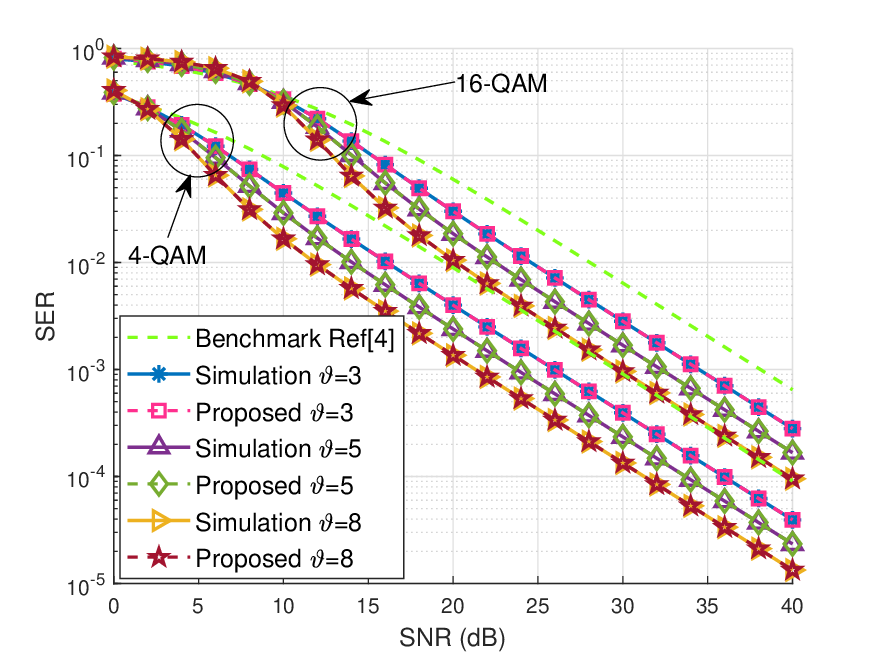}}
\caption{SER performance vs. SNR for 4-QAM and 16-QAM with ZF detection under Rayleigh fading. Theoretical vs numerical results for independent and correlated channel and noise.}%
 \label{fig:correlated_figures}
\end{figure*}


Therefore, $P_{e|(s_1,D_2)} \approx \frac{\sigma^2}{\pi}(I_3+I_4+I_5)$. Using the inherent symmetry of the integration domains, the average SER for independent channel and noise can be expressed as
\begin{align}
&P_{e|s_1} \notag\\
&\approx \frac{\sigma^2}{2\pi} \bigg[ 
   \frac{3\pi}{2\sigma^2} \left(1 - \frac{\gamma_1}{\sqrt{\sigma^2+\gamma_1^2}} \right)
   + \frac{\pi/2 - \arctan(\gamma_1/\sigma)}{\beta_1\sigma} - \frac{1}{2\gamma_1\beta_1} \notag\\
   &\quad + \frac{\arctan\left( \frac{\gamma_1}{\sqrt{\sigma^2+\beta_1^2}} \right) - \pi/2}{\beta_1\sqrt{\sigma^2+\beta_1^2}}
    + \frac{\beta_1^2+\gamma_1^2}{2\beta_1^2} \arctan\left(\frac{\beta_1}{\gamma_1}\right)
\bigg]. \label{SER_unc}
\end{align}
Equation~\eqref{SER_unc} offers a better approximation than equation~\eqref{SER_corr} for the uncorrelated case as can be observed in Fig.~\ref{fig-ind}.

\subsection{Copula Correlation Method}
When the joint distribution of channel fading and noise is not known, we model it using the copula method~\cite{nelsen2006introduction}. This approach is particularly useful for handling two random variables with different distributions. Using copulas, the joint PDF of \( n \) and \( h \) is given by
\begin{equation}
\label{eq:copulas}
 f_{|n|,|h|}(x,y) \! = \! C(F_{|n|}(x), F_{|h|}(y)) f_{|n|}(x) f_{|h|}(y),
\end{equation}
where $f_{|n|}(x)$, $f_{|h|}(y)$, and $F_{|n|}(x)$, $F_{|h|}(y)$ denote the marginal PDFs and CDFs of $|n|$ and $|h|$, respectively, and where \( C(\cdot,\cdot) \) denotes the bivariate copula density function. There are multiple families of copulas; in this work, we only consider the Frank method which  models strong dependencies between two variables~\cite{nelsen2006introduction},
\begin{equation}
    C(u, t)\!=\left( \frac{\!-\vartheta\!\left(e^{-\vartheta}\!-1\right)\ e^{-\vartheta(u + t)}}{\left( \left( e^{-\vartheta u} - 1 \right) \left( e^{-\vartheta t} - 1 \right) + \left( e^{-\vartheta} - 1 \right)\!\right)^2 } \right),\label{eq:frank}
\end{equation}
where \(\vartheta \!\in\! (-\infty, \infty) \!\setminus\! \{0\}\) governs dependence strength, with positive values capturing positive dependence. Using equation~\eqref{eq:copulas}, we calculate the PDF of 
the effective noise as
{\begin{align}
     f_{Z_r,Z_i}(z_r,z_r)=\frac{1}{2\pi\sigma_n^2\sigma_h^2}\!\!&\int_{0}^{+\infty}\!\!\!\!y^3 \exp\left( -\frac{y^2(z_i^2+z_r^2)}{2\sigma_n^2}-\frac{y^2}{2\sigma_h^2} \right)\notag \\ &\times C(F_{|n|}(\sqrt{z_i^2+z_r^2}\, y), F_{|h|}(y)) dy.
    \label{effectiv-copula}
\end{align}}
Finally, the SER can be computed using~\eqref{PE-16QAM} and~\eqref{eq:gen}.

\section{Results and Discussion}\label{sec:sim}
We numerically evaluate the SER of the linear (ZF) detector using Monte Carlo simulations for $N = 2\times10^6$ trials and compare it to the proposed theoretical results. We assume Rayleigh fading that is potentially correlated with the additive Gaussian noise, and normalized equiprobable $M$-QAM constellations for $M = 4$ and $M = 16$. The symbols \(\gamma, \beta\) are located at \(\{\pm1\}\) for the $4$-QAM case, and at \(\{\pm1, \pm3\}\) for the $16$-QAM. 
For the independent scenario ($\lambda=0$), Fig.~\ref{fig-ind} shows that the theoretical SER curve derived using~\eqref{Pc-eq-corr} matches exactly the simulation results and the benchmark~\cite[eq. 8.107]{4544952}. In addition, the asymptotic and approximation expressions given by equations~\eqref{SER_corr} and \eqref{SER_unc}, respectively, align perfectly with the benchmark and simulation result. 

For the correlated case, we simulated the two proposed approaches for the effective noise, as represented by~\eqref{pdf-corr} and~\eqref{effectiv-copula}. Figure~\ref{fig-corr} illustrates the SER using the PDF described in~\eqref{pdf-corr}. 
The simulation results align with the theoretical and asymptotic curves for different values of $\lambda$. Moreover, we observe in Fig.~\ref{fig-corr} that the benchmark calculations for the SER provide an upperbound on the ``correct" SER values under correlation, thus showing the overestimation in the SER values whenever correlation is not taken into consideration. For example, when $\abs{\lambda} \approx 1$, which represents a very high correlation, the performance deviates significantly from the independent case (benchmark curve), resulting in an overestimation of the SNR of approximately $16$~dB for $P_e=10^{-3}$ for both $4$ and $16$-QAM. We note that the standard errors of the Monte Carlo SER values are found to be within $[2\times 10^{-4}; 4.5 \times 10^{-6}]$. 


Similar conclusions are drawn from Fig.~\ref{fig-frank} for the Frank copula method. As observed in Figures~\ref{fig-corr} and~\ref{fig-frank}, the SER performance progressively deviates from the independent case as the correlation increases. This underscores the significant impact of channel-noise correlation on system performance. The observed behavior is attributed to the fact that, under statistical dependence, the knowledge of the channel $h$ reduces the uncertainty of the noise, which improves the detection accuracy. The proposed theoretical SER expressions accurately capture this behavior and closely match the simulation results, demonstrating the robustness of our method under both correlated and independent scenarios.

\section{Conclusion}\label{sec:conclusion}
In this paper, we propose and use a new approach to derive SER expressions for SISO systems under potentially correlated channel and noise variables, assuming ZF detection and $M$-QAM modulation in fast-fading Rayleigh channels. 
Our results show a strong agreement between numerical simulations and theoretical derivations. Furthermore, our framework highlights the detrimental effect of wrongly assuming independence between the channel and the additive noise, which can lead to an overestimation of the probability of detection error. The proposed approach is crucial for analyzing system performance under arbitrary channel and noise distributions and under different assumptions regarding channel-noise correlation, highlighting its general utility in assessing multiple emerging communication system paradigms.
Moving forward, we believe that accounting for physically consistent, real-world impairments and pursuing empirical validation are important directions for future research based on the theoretical framework introduced in this work.

\appendices
\label{sec:app}
\counterwithin*{equation}{section}
\renewcommand\theequation{\thesection.\arabic{equation}}

\bibliographystyle{IEEEtran}
\bibliography{IEEEabrv,my_bibliography}
\end{document}